\documentclass[12pt]{article}
\usepackage[dvips]{graphicx}

\begin{document}

\title{Intermediate state interaction in the particle oscillations in absorbing matter}
\author{V.I.Nazaruk\\
Institute for Nuclear Research of RAS, 60th October\\
Anniversary Prospect 7a, 117312 Moscow, Russia.*}

\date{}
\maketitle
\bigskip

\begin{abstract}
It is shown that the self-energy of intermediate particles plays a crucial role in the particle oscillations in absorbing matter.
\end{abstract}

\vspace{5mm}

\vspace{5mm}

\vspace{1cm}

*E-mail: nazaruk@inr.ru

\newpage
\setcounter{equation}{0}
\section{Introduction}
In the standard calculations of $ab$ oscillations in the medium [1,2] the one-particle (potential) model is used. The interaction of particles $a$ and $b$ with the matter is described by the potentials $U_{a,b}$. ${\rm Im}U_b$ is responsible for loss of $b$-particle intensity. In
particular, this model is used for the $ab$ transitions in a medium followed by absorption of $b$-particle:
\begin{equation}
(a-\mbox{medium})\rightarrow (b-\mbox{medium})\rightarrow f.
\end{equation} 
Here $(b-\mbox{medium})\rightarrow f$ represents the $b$-particle absorption or decay. It is common knowledge that the models with non-Hermitian operators have very restricted area of applicability. In [3-5] this problem has been studied by the example of $n\bar{n}$ transitions in a medium [6-14] followed by annihilation
\begin{equation}
(n-\mbox{medium})\rightarrow (\bar{n}-\mbox{medium})\rightarrow M,
\end{equation}
where $M$ are the annihilation mesons. The reason is that for the process (2) the absorption in the final state (annihilation) is extremely strong. This is simplest 2-step process involving intermediate-state interaction and absorption in the final state since the $n\bar{n}$ transition vertex corresponds to 2-tail diagram.  It was shown [3-5] that one-particle model mentioned above does not describe the process (2) as well as the total $n\bar{n}$ transition probability, because unitarity condition (optical theorem) is used for the essentially non-unitary $S$-matrix. The interaction Hamiltonian contains the antineutron optical potential $U_{\bar{n}}$ and ${\rm Im}U_{\bar{n}}$ plays a crucial role. The $S$-matrix should be {\em unitary}. Since the annihilation is the main effect which defines the speed of process (2) in the medium, the potential model should be rejected. (Potential model describes only $ab$ transitions with $b$-particle in the final state [3] when unitarity condition is not used.) 

In [4,5] the field-theoretical approach with final time interval has been proposed. However, the model with bare propagator (see Fig. 1a), that is, the case with antineutron self-energy $\Sigma =0$ has been considered only. In this paper the model with dressed propagator (see Fig. 1b) is studied as well. It is shown that the result is very sensitive to the $\Sigma $. The uncertainty in the value of the lower limit on the free-space $n\bar{n}$ oscillation time $\tau $ is conditioned by the uncertainty in the value of $\Sigma $.

The principal object of this paper is to study the role of intermediate state interaction, namely, the role of particle self-energy induced by intermediate state interaction. The conclusion obtained for process (2) is generalized to process (1). For example, the possible $\Lambda \bar{\Lambda }$ transitions [15,16] in a medium followed by $\bar{\Lambda }$ decay:
\begin{equation}
(\Lambda -\mbox{medium})\rightarrow (\bar{\Lambda }-\mbox{medium})\rightarrow (\pi ^0\bar{n} -\mbox{medium}).
\end{equation}

\section{Calculations}
The qualitative picture of the process (2) is as follows. The free-space $n\bar{n}$ transition comes from the exchange of Higgs bosons with the mass $m_H>10^5$ GeV [7] and so the subprocess of $n\bar{n}$ conversion is scarcely affected by a medium effects. From the dynamical point of view this is a momentary process: $\tau _c\sim 1/m_H<10^{-29}$ s. The antineutron annihilates in a time $\tau _a\sim 1/\Gamma $, where $\Gamma  $ is the annihilation width of $\bar{n}$ in the medium. We deal with two-step process with the characteristic time $\tau _{ch}\sim \tau _a$.

We consider the process (2). The neutron wave function is 
\begin{equation}
n(x)=\Omega ^{-1/2}\exp (-ipx).
\end{equation}
Here $p=(\epsilon ,{\bf p})$ is the neutron 4-momentum; $\epsilon ={\bf p}^2/2m+U_n$, where $U_n$ is the neutron potential. The interaction Hamiltonian has the form
\begin{equation}
H_I=H_{n\bar{n}}+H,
\end{equation}
\begin{equation}
H_{n\bar{n}}(t)=\int d^3x(\epsilon _{n\bar{n}}\bar{\Psi }_{\bar{n}}(x)\Psi _n(x)+H.c.),
\end{equation}
$H(t)=\int d^3x{\cal H}(x)$. Here $H$ is the Hamiltonian of the $\bar{n}$-medium interaction, $H_{n\bar{n}}$ is the Hamiltonian of $n\bar{n}$ conversion [10], $\epsilon _{n\bar{n}}$ is a small parameter with $\epsilon _{n\bar{n}}=1/\tau _{n\bar{n}}$, where $\tau _{n\bar{n}}$ is the free-space $n\bar{n}$ oscillation time ($\tau _{n\bar{n}}>\tau $ by definition); $m_n=m_{\bar{n}}=m$. 

In the lowest order in $H_{n\bar{n}}$ the amplitude of process (2) is {\em uniquely} determined by the Hamiltonian (5):
\begin{equation}
M=\epsilon _{n\bar{n}}G_0M_a,
\end{equation}
\begin{equation}
G_0=\frac{1}{\epsilon _{\bar{n}} -{\bf p}_{\bar{n}}^2/2m-U_n+i0},
\end{equation}
${\bf p}_{\bar{n}}={\bf p}$, $\epsilon _{\bar{n}}=\epsilon $. Here $G_0$ is the antineutron propagator. The corresponding diagram is shown in Fig. 1a. The amplitude of antineutron annihilation in the medium $M_a$ is given by 
\begin{equation}
<\!f0\!\mid T\exp (-i\int dx{\cal H}(x))-1\mid\!0\bar{n}_{p}\!>=N(2\pi )^4\delta ^4(p_f-p_i)M_a.
\end{equation}
Here $\mid\!0\bar{n}_{p}\!>$ is the state of the medium containing the $\bar{n}$ with the 4-momentum $p=(\epsilon ,{\bf p})$; $<\!f\!\mid $ denotes the annihilation products, $N$ includes the normalization factors of the wave functions. The antineutron annihilation width $\Gamma $ is expressed through $M_a$:
\begin{equation}
\Gamma \sim \int d\Phi \mid\!M_a\!\mid ^2.
\end{equation}

Since $M_a$ contains all the $\bar{n}$-medium interactions followed by annihilation including antineutron rescattering in the initial state, the antineutron propagator $G_0$ is bare. Once the antineutron annihilation amplitude is defined by (9), the expression for the process 
amplitude (7) {\em rigorously follows} from (5). For the time being we do not go into the 
singularity $G_0\sim 1/0$.

\begin{figure}[h1]
  {\includegraphics[height=.25\textheight]{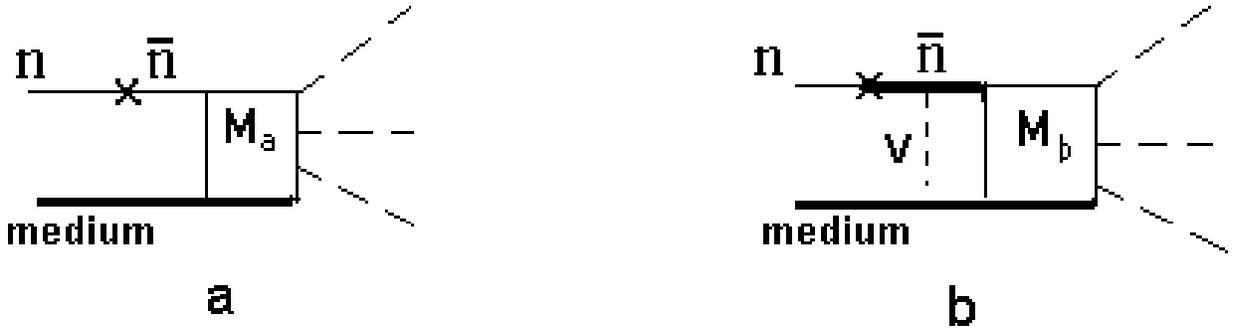}}
  \caption{{\bf a} $n\bar{n}$ transition in the medium followed by annihilation.  {\bf b} Same as {\bf a} but the antineutron
propagator is dressed (see text)}
\end{figure} 

One can construct the model with the dressed propagator. For the density of the Hamiltonian ${\cal H}$ we consider the model
\begin{eqnarray}
{\cal H}={\cal H}_a+V\bar{\Psi }_{\bar{n}}\Psi _{\bar{n}},\nonumber\\
H(t)=\int d^3x{\cal H}(x)=H_a(t)+V,
\end{eqnarray}
where ${\cal H}_a$ is the effective annihilation Hamiltonian in the second quantization representation, $V$ is the residual scalar field. 
We include the scalar field $V$ in the antineutron Green function
\begin{equation}
G_d=G_0+G_0VG_0+...=\frac{1}{(1/G_0)-V}=-\frac{1}{V}=-\frac{1}{\Sigma }.
\end{equation}
In line with physical meaning of Eq. (12) (power series in self-energy) we denoted $V=\Sigma $, where $\Sigma $ is the antineutron self-energy. The process amplitude is (see Fig. 1b)
\begin{equation}
M=\epsilon _{n\bar{n}}G_dM_b,
\end{equation}
$G_dM_b=G_0M_a$. The block $M_b=M_b({\cal H}_a,V)$ should describe the antineutron annihilation. Both $G_d$ and $M_b$ depend on $V$. Because of this the block (amplitude) $M_b$ cannot be naturally obtained from the formal expansion of the $T$-operator 
\begin{equation}
T\exp (-i\int dx({\cal H}_a+V\bar{\Psi }_{\bar{n}}\Psi _{\bar{n}})),
\end{equation}
as is the case for $M_a$ (see (9)). An additional restrictions on the expansion of (14) are required. 

The models shown in Figs. 1a and 1b we denote as the models {\bf a} and {\bf b}, respectively. If $\Sigma \rightarrow 0$, the model {\bf b} goes into model {\bf a}. In this sense the model {\bf a} is the limiting case of the model {\bf b}. We consider the model {\bf b}. For the process width $\Gamma _b$ one obtains
\begin{equation}
\Gamma _b=N_1\int d\Phi \mid\!M\!\mid ^2=\frac{\epsilon _{n\bar{n}}^2}{\Sigma ^2}N_1\int d\Phi
\mid\!M_b\!\mid ^2=\frac{\epsilon _{n\bar{n}}^2}{\Sigma ^2}\Gamma ',
\end{equation}
\begin{equation}
\Gamma '=N_1\int d\Phi \mid\!M_b\!\mid ^2,
\end{equation}
where  $\Gamma '$ is the annihilation width of $\bar{n}$ calculated through the $M_b$ (and not $M_a$). The normalization multiplier $N_1$ is the same for $\Gamma _b$ and $\Gamma '$. We recall the amplitude $M_b$ is unknown. (The antineutron annihilation width $\Gamma $ is expressed through the $M_a$.) For the estimation we put
\begin{equation}
M_b=M_a, \quad \Gamma '=\Gamma.
\end{equation}
This is an uncontrollable approximation. Now the process width is
\begin{equation}
\Gamma _b=\frac{\Gamma }{\tau _{n\bar{n}}^2\Sigma ^2}.
\end{equation}
The time-dependence is determined by the exponential decay law:
\begin{equation}
W_b(t)=1-e^{-\Gamma _bt}\approx \frac{\epsilon _{n\bar{n}}^2}{\Sigma ^2}\Gamma 't=\frac{\Gamma }{\tau _{n\bar{n}}^2\Sigma ^2} t.
\end{equation}
Equations (18) and (19) illustrate the result sensitivity to the value of parameter $\Sigma $. The physical meaning of $\Gamma _b$ and $W_b$ is standard: there is one-particle metastable state described by wave function (4). The width and probability of two-step decay (2) are given by (18) and (19), respectively.

Let $\tau ^b$ and  $T_{n\bar{n}}$ are the lower limit on the free-space $n\bar{n}$ oscillation time calculated by means of model {\bf b} and the experimental bound on the neutron lifetime in a nucleus [13], respectively. From the condition $W_b(T_{n\bar{n}})<1$ we have
\begin{equation}
\tau _{n\bar{n}}>\tau ^b=\frac{1}{\Sigma }\sqrt{\Gamma T_{n\bar{n}}}.
\end{equation}.
For estimation we take $\Gamma =100$ MeV and $\Sigma =10$ MeV. Using the value $T_{n\bar{n}}>1.77\cdot 10^{32}$ yr obtained by Super-Kamiokande collaboration [13], one obtains
\begin{equation}
\tau ^b=1.2\cdot 10^{9}\; {\rm s}.  
\end{equation}
This value exceeds the restriction given by the Grenoble reactor experiment [8] by a factor of 14 and the lower limit given by potential model [9-13] by a factor of 5.

At the point $\Sigma =0$ Eqs. (19) and (20) are inapplicable. In this connection we return to the model {\bf a}. The amplitude (7) diverges
\begin{equation}
M=\epsilon _{n\bar{n}}G_0M_a\sim \frac{1}{0},
\end{equation}
$G_0\sim 1/0$. These are infrared singularities conditioned by zero momentum transfer in the $n\bar{n}$ transition vertex. (In the model {\bf b} the effective momentum transfer $q_0=V=\Sigma $ takes place.) For solving the problem the field-theoretical approach with finite time interval [14] is used. It is infrared free. In the case of potential model, the approach with finite time interval reproduces all the results on the particle oscillations obtained by means of potential model (see sect. 5.2 of ref. [4]). This is the test of above-mentioned approach. However, our purpose is to describe the process (2) by means of Hermitian Hamiltonian because the potential model does not describe the absorption.

For the model {\bf a} the process (1) probability was found to be [4,5]
\begin{equation}
W_a(t)\approx  W_f(t)=\epsilon _{n\bar{n}}^2t^2, \quad \Gamma t\gg 1,
\end{equation}
where $W_f$ is the free-space $n\bar{n}$ transition probability. Owing to annihilation channel, $W_a$ is practically equal to the free-space $n\bar{n}$ transition probability. If $t\rightarrow \infty $, Eq. (23) diverges just as the modulus (22) squared does. If 
$\Sigma \rightarrow 0$, Eq. (18) diverges quadratically as well. Equation (23) does not contain density-dependence ($\Gamma-$dependence) since it corresponds to the limiting case $\Gamma t\gg 1$ which is realized for the $n\bar{n}$ transition in nuclei [4,5].

The explanation of the $t^2$-dependence is simple. The process shown in Fig. 1a represents two consecutive subprocesses. The speed and probability of the whole process are defined by those of the slower subprocess. If $1/\Gamma \ll t$, the annihilation can be considered 
instantaneous. So, the probability of process (2) is defined by the speed of the $n\bar{n}$ transition: $W_a\approx W_f\sim t^2$. 

Distribution (23) leads to very strong restriction on the free-space $n\bar{n}$ oscillation 
time:
\begin{equation}
\tau ^a=10^{16}\; {\rm yr}.
\end{equation}
Here $\tau ^a$ is the lower limit on the free-space $n\bar{n}$ oscillation time calculated by means of model {\bf a}. 

For the process (1) instead of (19) we have
\begin{equation}
W_b(t)\approx \frac{\Gamma _b}{\tau _{ab}^2\Sigma _b^2} t.
\end{equation}
Here $\tau _{ab}$ is the free-space $ab$ oscillation time, $\Sigma _b$ is the $b$-particle self-energy, $\Gamma _b$ is the absorption (decay) width of $b$-particle in the case of absorption (decay).

\section{Discussian and conclusion}
In this paper we don't touch the application of above-considered approach to the specific experiments on oscillations of ultra cold neutrons in storage vessels [17] and neutron disapearance into another braneworld [18]. Also we don't consider the calculations based on the diagram technique for direct reactions [19,20] since they are inapplicable to the problem under study [21]. (For example, the result $\Gamma \sim 1/E_n^2$ [19,20] ($E_n$ is the neutron binding energy) diverges when $E_n \rightarrow 0$.) Our goal is to study the role of intermediate state interaction.

As is seen from Eqs. (19) and (25), the self-energy plays the same part as the energy gap $U_{\bar{n}}-U_n$ in the potential model: it tends to suppress the processes (1) and (2). If $\Sigma \rightarrow 0$, $W_b$ rises quadratically. So $\tau ^b$ and $\tau ^a$ can be considered as the estimations from below and above, respectively. In both models the antineutron propagators don't contain the annihilation loops since the annihilation is taken into account in the amplitudes $M_a$ and $M_b$; the interaction Hamiltonians $H_I$ and unperturbed Hamiltonians are the same. The sole physical distinction between models {\bf a} and {\bf b} is the non-zero antineutron self-energy in the model {\bf b} which is conditioned by intermediate state interaction. However, it leads to the fundamentally different results (see (21) and (24)).
 
There is no rigorous recipes for calculation of $\Sigma $. One can adduce a great deal of arguments in support of the value $\Sigma =0$ [4, 22]. In particular, in contrast to the model {\bf b}, the model {\bf a} describes the following limiting cases [22]:

a) In the low-density approximation the model {\bf a} reproduces the branching ratio of channels of free-space $\bar{n}N$ interaction:
\begin{equation}
\frac{\Gamma _a}{\Gamma _t}= \frac{\sigma _a}{\sigma _t}. 
\end{equation}
Here $\sigma _a$ and $\sigma _t$ are the cross section of free-space $\bar{n}N$ annihilation and total cross section of free-space $\bar{n}N$ interaction, respectively; $\Gamma _a$ and $\Gamma _t$ are the width of process (2) and total width of $n\bar{n}$ transition in a medium.
($\Gamma _t$ includes the channels with antineutron and annihilation mesons in the final state.)

b) The model {\bf a} describes the process (2) on a free nucleon: 
\begin{equation}
n+N\rightarrow \bar{n}+N \rightarrow M.
\end{equation}
However, the small change of antineutron self-energy $\Sigma =0 \rightarrow \Sigma =V\neq 0$, or, similarly, effctive momentum transfer in the $n\bar{n}$ transition vertex converts the model {\bf a} to the model {\bf b}: $W_a\rightarrow W_b$ with $W_b\ll W_a$. This is because the process amplitude is in the peculiar point (see (7) and (8)) owing to zero momentum transfer. If $\Sigma $ changes in the limits $10\; {\rm MeV}>\Sigma >0\; {\rm MeV}$, the lower limit on the free-space $n\bar{n}$ oscillation time is in the range $10^{16}\; {\rm yr}>\tau >1.2\cdot 10^{9}\; {\rm s}$.

Finally, the problem is extremely sensitive to the $\Sigma $. This conclusion is also true for the process (1) since for this processes the $\Sigma $-dependence of the result is the same (see (25)). Further investigations are desirable. The self-energy of intermediate particle should be studied first.

\end{document}